\documentclass[12pt]{article}
\date{December, 2013}
\usepackage{mheck}

\institution{HARVARD}{\   Jefferson Physical Laboratory, Harvard University, Cambridge, MA 02138, USA}
\institution{IMPERIAL}{\ Theoretical Physics Group, Imperial College London, London, SW7 2AZ, UK}

\title{ Complete graphs, Hilbert Series, \\ and the Higgs Branch of the 4d $\cn=2$  $(A_n, A_m)$ SCFT's }
\authors{Michele Del Zotto \worksat{\HARVARD} \footnote{e-mail: {\tt eledelz@gmail.com}} and Amihay Hanany \worksat{\IMPERIAL} \footnote{e-mail: {\tt hanany@ic.ac.uk}} }

\abstract{ The strongly interacting 4d $\cn=2$ SCFT's of type $(A_n , A_m)$ are the simplest examples of models in the $(G,G^\prime)$ class introduced by Cecotti, Neitzke, and Vafa in \arXiv{1006.3435}. These systems have a known 3d $\cn=4$ mirror only if $h(A_n)$ divides $h(A_m)$, where $h$ is the Coxeter number. By 4d/2d correspondence, we show that in this  case these systems have a nontrivial global flavor symmetry group, and, therefore, a non-trivial Higgs branch. As an application of the methods of \arXiv{1309.2657}, we then compute the refined Hilbert series of the Coulomb branch of the 3d mirror for the simplest models in the series. This equals the refined Hilbert series of the Higgs branch of the $(A_n,A_m)$ SCFT, providing interesting information about the Higgs branch of these non--lagrangian theories.}

\begin{document}

\maketitle

\tableofcontents


\section{Introduction}

Let us denote by $A_n(X,Y) \equiv X^{n+1} + Y^2$ the $A_n$ singularity quasi--homogenous polynomial. The 4d $\cn=2$ models of type $(A_n,A_m)$ were obtained in \cite{CNV} by geometric engineering the Type II B superstring on the singular Calabi-Yau hypersurface of $\C^4$ defined as the zero locus of the polynomial\footnote{ In \cite{CNV} the more general class of systems $(G,G^\prime)$ for any pair of simple simply-laced Dynkin graphs was studied. The 3d mirrors of the more general $(G,G^\prime)$ models is still unknown.}
\be\label{AmAnge}
A_n(X_1,X_2) + A_m(Y_1,Y_2) = 0.
\ee
The LHS of eqn.\eqref{AmAnge} is a quasi--homogenous polynomial in four variables with weights
\be\label{AmAnweights}
q(X_1) = 1/h(A_n) \quad q(X_2) = 1/2 \quad q(Y_1) = 1/h(A_m) \quad q(Y_2) = 1/2.
\ee
As the sum of the weights in eqn.\eqref{AmAnweights} is bigger than one, by the criterion of \cite{GVW}, the singularity in eqn.\eqref{AmAnge} is at a finite distance in CY moduli space, and the model defines an honest 4d $\cn=2$ SCFT. Deforming eqn.\eqref{AmAnge} by lower order terms, we resolve the singularity, and the theory flows along its Coulomb branch. The properties and BPS spectra of the Coulomb branch of these systems were studied extensively in \cite{CNV} by 4d/2d worldsheet/target correspondence and BPS quivers. The Higgs branch of the moduli space remains mysterious and is the subject of the present letter.

\bigskip

By dimensional reduction on a circle, a 4d $\cn=2$ model reduces to a 3d $\cn=4$ system. In many cases, the 3d $\cn=4$ theories obtained from strongly-interacting 4d $\cn=2$ superconformal models in this way are non-lagrangian, but nevertheless have \emph{lagrangian} 3d mirrors \cite{BeniniXie,XieNano}. The Higgs and Coulomb branches of the infrared moduli spaces of 3d mirror systems are both hyper--K\"ahler and get exchanged under 3d mirror symmetry \cite{IntSei}. Clearly one can use the known lagrangian structures to get interesting informations about the non--lagrangian ones, and whether some of the information about the 3d $\cn=4$ system can be lifted to the four dimensional one is a very interesting question that was answered long time ago \cite{SW43}: for 3d-4d systems with 8 supercharges related by dimensional reduction on a circle, the 3d Coulomb branch is an elliptic fibration over the 4d Coulomb moduli, while the 3d and 4d Higgs branches \emph{coincide}. The Coulomb branch of its lagrangian 3d mirror, therefore, \emph{equals}, as a hyper--K\"ahler manifold, the Higgs branch of the parent 4d $\cn=2$ non--lagrangian system.

\bigskip

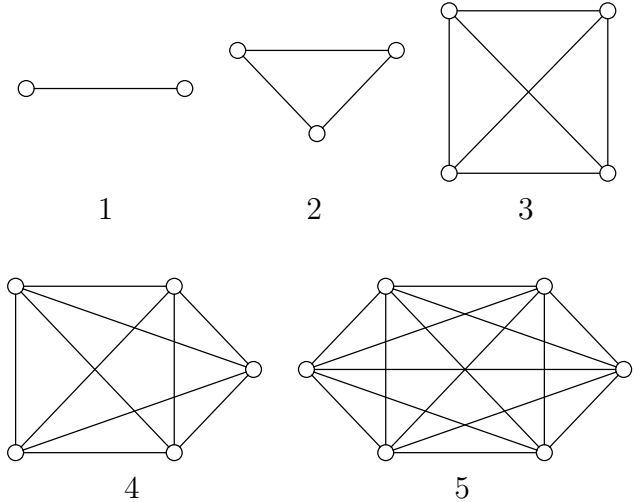
\begin{figure}
\begin{center}
\begin{tabular}{ccc}
$\xymatrix{*+[o][F-]{} \ar@{-}[rr] && *+[o][F-]{} }$ &
$ \begin{gathered}
\xymatrix{*+[o][F-]{} \ar@{-}[rr]\ar@{-}[dr] && *+[o][F]{} \\
&*+[o][F-]{}\ar@{-}[ur] & }
\end{gathered}$&$\begin{gathered}
\xymatrix{*+[o][F-]{} \ar@{-}[dd]\ar@{-}[rr]\ar@{-}[ddrr] && *+[o][F-]{}\ar@{-}[dd]\ar@{-}[ddll] \\
&&\\
*+[o][F-]{}\ar@{-}[rr] && *+[o][F-]{}}
\end{gathered}$\\
$ 1$& $2$ & $3$\\
\end{tabular}

\bigskip

\begin{tabular}{cc}
$\begin{gathered}
\xymatrix{*+[o][F-]{} \ar@{-}[dd]\ar@{-}[rr]\ar@{-}[ddrr] && *+[o][F-]{}\ar@{-}[dd]\ar@{-}[ddll] & \\
&&&*+[o][F-]{}\ar@{-}[ul]\ar@{-}[ulll]\ar@{-}[dl]\ar@{-}[dlll]\\
*+[o][F-]{}\ar@{-}[rr] && *+[o][F-]{}&}
\end{gathered}$&$\begin{gathered}
\xymatrix{&*+[o][F-]{} \ar@{-}[dd]\ar@{-}[rr]\ar@{-}[ddrr] && *+[o][F-]{}\ar@{-}[dd]\ar@{-}[ddll] & \\
*+[o][F-]{}\ar@{-}[ur]\ar@{-}[urrr]\ar@{-}[dr]\ar@{-}[drrr]\ar@{-}[rrrr]&&&&*+[o][F-]{}\ar@{-}[ul]\ar@{-}[ulll]\ar@{-}[dl]\ar@{-}[dlll]\\
&*+[o][F-]{}\ar@{-}[rr] && *+[o][F-]{}&}\end{gathered}$\\
$4$ & $5$ \\
\end{tabular}
\end{center}
\caption{Some of the quivers for the 3d $\cn=4$ mirrors of the models of type $\ca_{s,p}$: $s$ is the number indicated under the quivers, and each edge has multeplicity $p$. Notice that these graphs are all $\mathfrak{S}_{s+1}$ symmetric under permutation of the nodes, and that removing one node maps the 3d $\ca_{s,p}$ quiver into the $\ca_{s-1,p}$ one.}\label{3dmirrors}
\end{figure}

Indeed, the 3d $\cn=4$ mirrors of some of the dimensionally reduced $(A_n,A_m)$ models were found in \cite{Xie} and turn out to be lagrangian $3d$ $\cn=4$ systems.\footnote{ The result heavily builds on previous work on Hitchin systems in \cite{Boalch}.} When $h(A_n)$ divides $h(A_m)$ 3d mirrors are known: In section \ref{AnAmfl}, we show that in this case the $(A_n,A_m)$ systems can have a non--trivial flavor symmetry, and therefore the existence of a 3d $\cn=4$ mirror is expected. The $(A_n,A_m)$ systems such that $h(A_n)$ divides $h(A_m)$ are the 4d $\cn=2$ models that we consider in this paper, and we find convenient to introduce the following notation for this subclass
\be
\framebox{$\ca_{s,p} \equiv (A_s,A_{(s+1)p-1})$}
\ee
The rank of the flavor group for such a system is $s$. The 3d $\cn=4$ mirror of a system of type $\ca_{s,p}$, is an abelian $\cn=4$ quiver gauge theory. The underling graph has $s+1$ nodes and $p$ edges in between each pair of nodes --- see figure \ref{3dmirrors}. To each node corresponds a $U(1)$ gauge group, while edges denotes as usual charged bifundamental $\cn=4$ hypermultiplets. By abuse of notation, we are going to denote these graphs with the symbol $\ca_{s,p}$ too.\footnote{ Beware! The graph $\ca_{s,p}$ has $s+1$ nodes! Indeed, there is an overall $U(1)$ which decouples, and the gauge group is $U(1)^{s+1}/U(1)$.} The graph $\ca_{s,1}$ is known as the \emph{complete graph} of type $\ck_{s+1}$ in graph theory. 

\bigskip

The 4d Higgs branch of the $\ca_{s,p}$ system, is determined by the Coulomb branch of its 3d mirror. As opposed to the Higgs branch, that has no quantum corrections (hyper--K\"ahler quotient construction \cite{HKLR}) and is determined classicaly, the Coulomb branch of a 3d $\cn=4$ system receives both perturbative 1-loop corrections, and infinitely many non--perturbative istantonic corrections. Even if the model is lagrangian, it seems hopeless to be able to compute it for the quiver theories in figure \ref{3dmirrors}. Very recently, however, a novel approach to the characterization of the Coulomb branch of a 3d $\cn=4$ system was proposed in \cite{CHZ}. The essential idea is that the quantum corrections to the Coulomb branch can be characterized in terms of GNO dual \cite{GNO} BPS monopole operators.\footnote{Notice that all the theories that we are going to consider satisfy the unitarity bound $\Delta \geq 1/2$ of \cite{GW3d}.}  In reference \cite{CHZ}  a general formula for Hilbert series of the Coulomb branch of a 3d $\cn=4$ unitary theory is derived. Such Hilbert series enumerates gauge invariant operators modulo chiral ring relations, and is expressed only in terms of the BPS monopole operator charges suitably `dressed' to keep track of the residual classical gauge degrees of freedom. The algebraic structure of the Coulomb branch of the 3d mirrors of $\ca_{s,p}$ models is captured by the plethystic logarithm of the Hilbert series of \cite{CHZ}. Even if the approach does not allow one to compute the exact quantum corrected hyper--K\"ahler metric on the Coulomb branch, in some  circumstances it is powerful enough to predict explicitly the structure of the full quantum moduli space as an algebraic variety. Using this method we will study the 4d Higgs branches of the elements of the $\ca_{s,p}$ class for $s=1,2,3$ and arbitrary $p$, and completely determine their structures as algebraic varieties.\footnote{ It is well--known that the Hilbert series of the Higgs branch of a 4d $\cn=2$ model corresponds, under favorable circumstances, to the Hall--Littlewood limit of the superconformal index \cite{GRsci}, the latter being unknown even for the simpler $ADE$ Argyres--Douglas systems.}

\bigskip

The rest of the paper is organized as follows: in section \ref{AnAmfl} we compute the rank of the flavor symmetry group of a model of type $(A_n,A_m)$; in section 3 we give a brief review about Hilbert series and plethysm; in section \ref{HilbSer} we specialize the general formula of \cite{CHZ} to our case, and discuss some universal features of the Higgs branches of the class of models $\ca_{s,p}$ ; we conclude by computing explicitly the Hilbert series of the models of type $\ca_{s,p}$, for $s=1,2,3$, and by determining the algebraic structure of the corresponding Higgs branches.

\section{The rank of the flavor group of the $(A_n,A_m)$ models}\label{AnAmfl}

The models $(A_n,A_m)$ have, by construction, the BPS quiver property \cite{CV11,CNV}. For four--dimensional $\cn=2$ theories with BPS quivers the rank of their global symmetry group is determined by 4d/2d target/worldsheet correspondence with the same method used in \cite{Arnold1,Arnold2}. The charge lattice $\Gamma$ of quantized electric, magnetic and flavor charges of the model has a symplectic structure induced by the Dirac electromagnetic antisymmetric pairing. By definition, the flavor charges of $\Gamma$ belongs to the radical of the Dirac pairing, $i.e.$ if we denote by $B$ the integer antisymmetric matrix associated to the Dirac pairing, the rank of the flavor group is equal to the dimension of the kernel of $B$. By 4d/2d correspondence, to any 4d $\cn=2$ theory corresponds a parent 2d $(2,2)$ model with $\hat{c}<2$ that has the same BPS quiver. Let us denote by $S$ the $tt^*$ Stokes matrix of the 2d $(2,2)$ model \cite{CV92}. We have
\be
B = S^t - S.
\ee
If $f\in \Gamma$ is a flavor charge, $B f = 0$, and therefore the rank of the flavor group equals the number of $+1$ eigenvalues of the matrix $(S^{-1})^t S$, $i.e.$ the 2d monodromy matrix of \cite{CV92}. One of the main properties of the 2d monodromy matrix is that
\be
\text{Eigenvalues} \, (S^t)^{-1} S = \big\{ \text{exp}(2 \pi i \, q_{RR}( \Psi)), \quad \Psi \in \mathscr{R}\big\}
\ee
where $\mathscr{R}$ is the chiral ring of primary operators of the 2d model and $q_{RR}$ denotes Ramond-Ramond charge. Consider the 2d model with superpotential $A_n(X_1,X_2) + A_m(Y_1,Y_2)$: being a direct sum, the Hilbert space is a tensor product, and the corresponding Stokes matrix factorizes as $S = S_n \otimes S_m$. Then by eqn.\eqref{AmAnweights} the rank of the flavor group is given by the number of solutions of the following equations:
\be\label{AmAnflav}
\frac{\ell}{h(A_n)}+\frac{k}{h(A_m)} \in \mathbb{Z} \qquad \begin{aligned} & \ell = 1, \dots, n\\ & k = 1, \dots, m\end{aligned}
\ee
Since $(A_n,A_m)$ and $(A_m,A_n)$ denote the same theory by definition, let us choose $n\leq m$. Then eqn.\eqref{AmAnflav} has precisely $\text{gcd}[h(A_n),h(A_m)] - 1$ solutions. This shows that the rank of the flavor group of the $\ca_{s,p}$ model is $s$. It would be interesting to study the Higgs branches of the other models that have a nontrivial flavor symmetry: we leave this for future work \cite{DZH}.

\section{3d Coulomb branches, and Hilbert series plethysm}

To fix notation, let us summarize briefly here the findings of \cite{CHZ,FHH}. For a 3d $\cn=4$ system there are 2 types of supermultiplets: the vectormultiplet (consisting of a gauge field and 3 real scalars + fermions), and the hypermultiplet (consisting of 4 real scalars + fermions). Giving non--zero vevs (only) to the vectormultiplets (resp. hypermultiplets) we obtain the Coulomb (resp. Higgs) branch of the moduli space, that are exchanged under 3d mirror symmetry. As we have already mentioned in the introduction, while the Higgs branch is exact classicaly, the Coulomb branch receives all sorts of quantum corrections. Recall that the vectors in 3d are dual to scalars. The bosonic part of an abelian vectormultiplet is equivalently characterized in terms of 4 real scalars. Instead of working in the field basis defined by the modes of these 4 real scalars, we group two of them in a complex $\cn=2$ scalar, $\phi$, and replace the modes associated to the other two fields by their electro--magnetic dual monopole operator counterparts \cite{BKW,BK}. This construction fixes implicitly a $\cn=2$ subalgebra of the 3d $\cn=4$ algebra by the choice for the field $\phi$. The BPS monopole operators and the modes of the field $\phi$ generate the $\cn=2$ chiral ring of the model on its Coulomb branch. The Hilbert series of the Coulomb branch, is essentialy, the Poincar\'e series of the chiral ring, where each chiral operator is weighted in terms of its IR conformal dimension $\Delta$. Notice that the complex adjoint scalar $\phi$ is not canonically normalized: $\Delta(\phi)=1$.\footnote{ As opposed to the usual case: in 3d scalars have $\Delta=1/2$.} Each monopole operator is characterized by $r$ topological GNO magnetic charges $(m_1,\dots,m_r)$, where $r$ is the rank of the gauge group\cite{GNO}: these charges determine the IR conformal dimensions of the monopole operator. The structure of the Hilbert series is roughly the following
\be
H = \sum_{\text{chiral ring}} t^{\Delta}
\ee
All the theories that we are going to consider in this work are abelian: this is a great simplification. For abelian theories there are no constraints on the possible GNO topological charges, these are simply integer valued. Moreover, $\phi$ is simply a complex scalar in this case, and therefore its contribution can be easily factored out in terms of a dressing factor given by $(1-t)^{-r}$. Then we can write the Hilbert series in a form that is more concrete
\be
H(t) = \frac{1}{(1-t)^r} \sum_{m_1,\dots,m_r \in \mathbb{Z}} t^{\Delta(m_1,\dots,m_r)}
\ee
where the sum is taken over all charges of all BPS monopole operators, weighted by the appropriate dimension. Such a series can be refined by introducing fugacities for the magnetic charges as follows
\be
H(t\,;z_1,\dots,z_r) = \frac{1}{(1-t)^r} \sum_{m_1,\dots,m_r \in \mathbb{Z}} t^{\Delta(m_1,\dots,m_r)} \prod_{i=1}^r (z_i)^{m_i}.
\ee
The relations of the chiral ring of operators are encoded in the multiplicities of the Hilbert series: the structure of the Coulomb moduli space as an algebraic variety is encoded in the plethystic logarithm of this function, defined as follows:
\be
\text{PL}[H(t\,;z_1,\dots,z_r)] \equiv \sum_{k\geq 1} \mu(k)\frac{\text{Log}[H(t^k,\,;z_1^k,\dots,z_r^k)]}{k}
\ee 
where $\mu(k)$ is the M\"obius function. Here are some of its fundamental properties \cite{FHH}:
\begin{itemize}
\item If the moduli space is a complete intersection variety the plethystic logarithm of the Hilbert series is a polynomial of finite degree, if not the plethystic Log grows indefinitely;
\item Multiplicities and charges of the generators of the chiral ring appears in the expansion of the plethystic logarithm as positive contributions;
\item Relations in between operators of given charges (defining the algebraic structure of the moduli) are encoded in the negative contributions to the plethystic Log.
\end{itemize}
Clearly, if the moduli space is not a complete intersection variety the plethystic Log is a series too. However, most of the negative contributions are redundant: these represent Hilbert syzygies in between relations. Typically, the moduli space is captured by the content of the first few negative contributions of the plethystic Log expansion. 

\medskip

In particular, the dimension of the moduli space is captured by the order of the pole of the Hilbert series at $t=1$.

\section{The refined Hilbert series for the $\ca_{s,p}$ models}\label{HilbSer}

\begin{figure}
\begin{center}
\begin{tabular}{cccc}
$\begin{gathered}\xymatrix{*+[F-]{}\\*+[o][F-]{} \ar@{-}[u]^{k}}\end{gathered}$&  &$\begin{gathered}\xymatrix{*+[F-]{}&&*+[F-]{}\\*+[o][F-]{} \ar@{-}[u]^{k}\ar@{-}[rr]^p&&*+[o][F-]{} \ar@{-}[u]_{k}}\end{gathered}$&$\begin{gathered}\xymatrix{*+[F-]{}&&*+[F-]{}\\*+[o][F-]{} \ar@{-}[u]^{k}\ar@{-}[rr]^p&&*+[o][F-]{} \ar@{-}[u]_{k}\\&*+[o][F-]{}\ar@{-}[ur]_p\ar@{-}[ul]^p&\\&*+[F-]{} \ar@{-}[u]^{k}&\\}\end{gathered}$\\
\end{tabular}
\end{center}
\caption{Examples of graphs of the class $\ca^{k}_{s,p}$ for $s = 1,2,3$.}\label{3dmirrors*}
\end{figure}
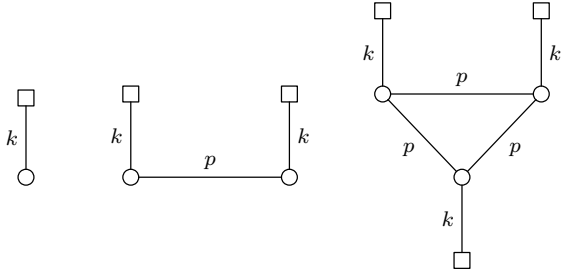

Consider the graph of type $\ca_{s,p}$ we have defined in the introduction. For the study of the refined Hilbert series of the Coulomb branch of these models, it is very useful to introduce another class of models, obtained from the one we are considering by, say, `adding flavors'. Let $\ca^{k}_{s,p}$ be the graph obtained from $\ca_{s-1,p}$ by adding $k$ flavors for each node (see figure \ref{3dmirrors*} for some examples). Each graph $\ca^{k}_{s,p}$ defines a $U(1)^s$ abelian 3d $\cn=4$ theory in the standard way. Such a theory has $s$ GNO integer valued magnetic charges, and the IR conformal dimension of a BPS monopole operator is 
\be\label{thedims}
\Delta(m_1,\dots,m_s) \equiv  \frac{p}{2} \left(\sum_{1\leq a < b \leq s} \mid m_a - m_b \mid\right) + \frac{k}{2}\left(\sum_{j=1}^s \mid m_j \mid\right)
\ee
This definition is motivated by the fact that the graph $\ca_{s,p}^p$ captures the physics of the 3d mirrors of the 4d systems of type $\ca_{s,p}$: Indeed,  since the graphs $\ca_{s,p}$ are connected, the gauge group that corresponds to these systems has a redundant $U(1)$, and getting rid of it we get precisely the model $\ca_{s,p}^p$. To see this, consider the IR conformal dimensions of the BPS monopole operators of the system. Since the graph $\ca_{s,p}$ has $s+1$ nodes, there are $s+1$ GNO magnetic charges $m_1,\dots,m_{s+1}$, that are not constrained in any ways, the gauge groups being abelian --- \emph{i.e.} these charges are integer valued. Moreover, the abelian nature of the quiver entails that the IR conformal dimension $\Delta$ of a BPS operator is determined directly from the edges of $\ca_{s,p}$ and from its GNO charges. We obtain
\be
\Delta(m_1,\dots,m_{s+1}) \equiv \frac{p}{2} \sum_{1\leq a < b \leq s+1} \mid m_a - m_b \mid
\ee
We can translate all magnetic charges simultaneously without altering $\Delta$, therefore we can set to zero one of them, all choices being equivalent by symmetry. We choose to set to zero $m_{s+1}$. The resulting IR dimension is
\be
\Delta(m_1,\dots,m_s) =  \frac{p}{2} \left(\sum_{1\leq a < b \leq s} \mid m_a - m_b \mid + \sum_{j=1}^s \mid m_j \mid\right)
\ee
That is the charge corresponding to the graph $\ca^p_{s,p}$. Therefore, we have $s$ classical complex scalar operators in the sense of \cite{CHZ}, and the dressing factor of the Hilbert series is $(1-t)^{-s}$. The same dressing factor, of course, is given for the model $\ca_{s,p}^k$.

Notice that all theories satisfy the unitarity bound $\Delta \geq 1/2$.

\bigskip

Having determined the structure of the GNO charges, the IR dimensions of the BPS monopole operators, and the dressing factor, the general formula of \cite{CHZ} for the refined Hilbert series of the Coulomb branch of the 3d mirrors of the $\ca_{s,p}$ models gives
\be\label{generalnaseria}
H_{s,p}(t\, ; z_1,\dots,z_s) \equiv \frac{1}{(1-t)^s} \sum_{m_1, \dots, m_s \in \mathbb{Z}} t^{\frac{p}{2} \left( \sum_{1\leq a < b \leq s} \mid m_a - m_b \mid  + \sum_{j=1}^s \mid m_j \mid\right)}\prod_{k=1}^s z_k^{m_k}
\ee
This formula has the structure
\be
H_{s,p}(t \, ; z_1,\dots,z_s) = (1-t)^{-s} f_s(t^p \, ; z_1,\dots,z_s)
\ee
where $f_s$ is a function of $s+1$ variables whose structure does not depend on $p$. Therefore we obtain the relation
\be\label{piden}
\framebox{$H_{s,p}(t\, ; z_1,\dots,z_s) = \left(\frac{1-t^p}{1-t}\right)^s \, H_{s,1}(t^p \, ; z_1,\dots,z_s)$}
\ee
Computing $H_{s,1}(t\, ;z_1,\dots,z_s)$ for a given $s$ fixes all the refined Hilbert series for all the elements of the family $\ca_{s,p}$ with the same $s$ and higher $p$'s: The combinatorics of the Hilbert series of these models is all encoded in the complete graph $\ck_s$, dressed with flavors, one per each node.

\subsection{Rewriting the Hilbert Series $H_{s,1}(t\,;z_1,\dots,z_s)$}

Let us denote by $H^{(k)}_{s}(t\, ; z_1,\dots,z_s)$ the refined Hilbert series associated to the graph $\ca^k_{s,1}$. Notice that, getting rid of the redundant $U(1)$ in the gauge group, we have the identity $H_{s,1} = H^{(1)}_s$. The function $H^{(k)}_{s}(t\, ; z_1,\dots,z_s)$ being a symmetric function of all of the $z_i$'s has the following structure (the constant term is one: this is the contribution form the identity operator in the $\cn=2$ chiral ring)
\bea
H^{(k)}_{s}(t \,; z_1,\dots,z_s) &=\frac{1}{(1-t)^s} \Big( 1 + \sum_{i=1}^s g_{k,s}^{(1)}(t,z_i) + \sum_{1\leq i_1 < i_2 \leq s} g_{k,s}^{(2)}(t,z_{i_1},z_{i_2})+\cdots\\
& \cdots + \sum_{1\leq i_1 < i_2 < \cdots < i_{s-1} \leq s} g_{k,s}^{(s-1)}(t,z_{i_1},\dots, z_{i_{s-1}}) + g_{k,s}^{(s)}(t,z_1,\dots,z_s)\Big)
\eea
where the $g_{k,s}^{(n)}$ are symmetric functions of their $n$ $z_i$'s variables; these functions have a clear physical meaning: the $g_{k,s}^{(n)}(t ; z_{i_1},\dots,z_{i_n})$ function represent the contribution to the refined Hilbert series $H^{(k)}_s$ from the sectors with non--vanishing $m_{i_1},\dots,m_{i_n}$ charges.  In particular we will see that the functions $g_{k,s}^{(s)}(t\,;z_1,\dots,z_n)$ play a special r\^ole, being equal to the contribution from the sector with all non--vanishing magnetic charges. It is natural to define
\be
\framebox{ $ h^{(s)}_k(t\,;z_1,\dots,z_n) \equiv g_{k,s}^{(s)}(t\,;z_1,\dots,z_n)$ }
\ee
From the structure of the graphs $\ca^k_{s,1}$ and of the IR conformal dimensions eqn.\eqref{thedims}, it is clear that
\be
g_{k,s}^{(n)}(t ; z_{i_1},\dots,z_{i_n}) = h^{(n)}_{k+s-n}(t ; z_{i_1},\dots,z_{i_n}),
\ee
\emph{i.e.}, each of these contribution can be accounted effectively as coming from a parent graph, obtained by treating as flavor nodes all the nodes associated to a  zero magnetic charge. Applying this reasoning to the Hilbert series for the $\ca_{s,1}$ systems ($k=1$ case above), we obtain
\bea
H_{s,1}(t\,;z_1,\dots,z_s) &= \frac{1}{(1-t)^s}\Big(1 + \sum_{i=1}^s h^{(1)}_s(t\,;z_i) +  \sum_{1\leq i_1 < i_2 \leq s} h_{s-1}^{(2)}(t,z_{i_1},z_{i_2})+\cdots\\
&\cdots + \sum_{1\leq i_1 < i_2 < \cdots < i_{s-1} \leq s} h_{2}^{(s-1)}(t,z_{i_1},\dots, z_{i_{s-1}}) + h_1^{(s)}(t,z_1,\dots,z_s)\Big)
\eea
In particular, the unrefined Hilbert series reads
\be
H_{s,1}(t) =\frac{1}{(1-t)^s}\Big( 1 + \sum_{j=1}^s \binom{s}{j} h^{(j)}_{s-j+1}(t)\Big)
\ee
These obvious remarks greatly simplify the computation of the Hilbert series for the family of models $\ca_{s,1}$.  Notice that the refined Hilbert series of the model $\ca_{s,p}$ is completely captured by the $h^{(n)}_k$ functions, that are, in a sense, fixed by the complete graph $\ck_n$.

\bigskip

As an application of the method we discussed, we are going to evaluate below the refined Hilbert series for the models $\ca_{s,p}$ for $s=1,2,3$.



\subsection{The case of $A_{2p-1}$ Argyres-Douglas theory}
$A_n$ Argyres--Douglas theories are identified with the elements $(A_1,A_n)$ of the series. Models $\ca_{1,p}$, corresponds to $A_{2p-1}$ Argyres--Douglas theories. If this is the case, the 4d $\cn=2$ Higgs branch is known to be $\C^2/\mathbb{Z}_p$  \cite{AMT}. Our result of section \ref{AnAmfl} reproduces the known fact that $A_{2k}$ Argyres--Douglas theories do not have a Higgs branch. The 3d $\cn=4$ mirror of the models $\ca_{1,p}$ is simply
\be
\xymatrix{*+[o][F-]{} \ar@{-}[rr]^p && *+[o][F-]{}}
\ee
Getting rid of the redundant $U(1)$ factor, this system is equivalent to $U(1)$ coupled to $p$ fundamentals
\be
 \xymatrix{*+[o][F-]{} \ar@{-}[rr]^p && *+[F-]{} }
\ee
The latter has a well-known Coulomb branch: it is precisely $\C^2/\mathbb{Z}_p$. Let us reproduce the same result using the formulas of \cite{CHZ}. The magnetic charge is
\be
\Delta = \frac{p}{2} |m|
\ee
The function $h^{(1)}_p(t\,;z )$ is given by the sum of the series
\be\label{h1p}
h^{(1)}_p(t\,;z ) = \sum_{m>0} (z^m + z^{-m}) t^{ \frac{p}{2} m }= \frac{t^{\frac{p}{2}}(z^2 - 2 \, t^{\frac{p}{2}} \, z+1)}{z(1-z\,t^{\frac{p}{2}})(1-\frac{1}{z} \, t^{\frac{p}{2}})}
\ee
The dressing factor is $(1-t)^{-1}$ from the $U(1)$ gauge group. The resulting refined series for $p=1$ is
\bea
H_{1,1}(t\, ; z) &\equiv \frac{1}{1-t}\Big( 1 + h^{(1)}_1(t\,;z ) \Big)\\
&= \frac{ 1 }{(1-z\,t^{\frac{1}{2}})(1-\frac{1}{z} \, t^{\frac{1}{2}})},
\eea
consistent with the observation that the Coulomb branch of the 3d $\cn=4$ model of $U(1)$ with one flavor is $\C^2$. For $p>1$, we have, from eqn.\eqref{piden}
\be\label{A1pHilb}
H_{1,p}(t\, ; z) = \left(\frac{1-t^p}{1-t}\right) \frac{ 1 }{(1-z\,t^{\frac{p}{2}})(1-\frac{1}{z} \, t^{\frac{p}{2}})}
\ee
In perfect agreement with (3.2) of \cite{CHZ}. Using eqn.\eqref{piden} The plethystic logarithm of $H_{1,p}(t,z)$ is given by
\be\label{sgagsg}
t + \left(z + \frac{1}{z}\right) t^{p/2} - t^p.
\ee
Notice that the plethystic logarithm is a polynomial: this means that the Higgs branch is a complete intersection. In all other cases that we are going to consider, this is not the case. From the plethystic Log we see that there is one complex scalar operator with dimension 1, $\Phi$, and two charged operators in dimension $p/2$. These are clearly the BPS monopole operators $V_{+1}$ and its PCT conjugate $V_{-1}$. In dimension $p$ we have one scalar relation. There is only one possible relation we can build in between these objects that has the right properties, namely $V_{+1}V_{-1} = \Phi^p$, that is precisely the equation of the $A_{p-1}$ singularity, $C^2/\mathbb{Z}_p$. 

\medskip

The refined Hilbert series $H_{1,p}(t\,; z)$ can be rewritten in terms of $SU(2)$ characters: One can rewrite eqn.\eqref{A1pHilb} as follows:\footnote{ Here we are using the simple root basis of the charge lattice, as opposed to the fundamental weight basis one.\label{caveat}}
\be
H_{1,p}(t\,;z) =\frac{1-t^p}{1-t} \sum_{m=0}^{\infty} [m] t^{pm/2}  
\ee
where $[m]$ is the character of the $SU(2)$ representation with Dynkin label $m$. Increasing $p$, the moduli space develops an orbifold singularity. Let us remark that even though there is an expansion of $H_{1,p}(t\,;z)$ in terms of $SU(2)$ characters, there is no $SU(2)$ flavor symmetry except for $p=2$. Indeed, if such a symmetry is there, we should find at $\Delta = 1$ the superpartners of the $SU(2)$ flavor currents, and from eqn.\eqref{sgagsg} there is an adjoint representation at $\Delta =1$ only for $p=2$. This is consistent with the fact that $\ca_{1,2}$ corresponds to the 4d $\cn=2$ $A_3$ Argyres-Douglas model, for which the rank two hidden symmetry enhances to $SU(2)$.

\subsection{The 4d Higgs branch of $\ca_{2,p}$ systems.}
The 3d mirror of the $\ca_{2,p}$ systems is given by
\be
\begin{gathered}
\xymatrix{*+[o][F-]{} \ar@{-}[rr]^p\ar@{-}[dr]_p && *+[o][F-]{} \\
&*+[o][F-]{}\ar@{-}[ur]_p & }
\end{gathered}
\ee
Notice that in the $p=1$ case, this graph reduces to the one for the reduced moduli space of one $SU(3)$ instanton on $\C^2$, $\widetilde{\cm}_{1,3}$ \cite{KN,IntSei,BHM}. This is precisely the case that we need to analyze. In view of eqn.\eqref{piden}, the refined Hilbert series $H_{2,p}$ for $p>1$ will be completely determined by the value of $H_{2,1}$. The IR conformal dimentions of the allowed magnetic monopoles are
\be
\Delta = \frac{1}{2}\Big(|m_1 - m_2| + |m_2 - m_3 | + |m_1 - m_3|\Big)
\ee
Using the $\mathbb{Z}$ translational symmetry $m_i \to m_i + n$, we can always set one magnetic charge to be zero, and the IR conformal dimention reduces to 
\be
\Delta(m_1,m_2) = \frac{1}{2}\Big(|m_1 - m_2| + |m_2  | + |m_1|\Big)
\ee
Indeed, the system is equivalent to
\be
\begin{gathered}
\xymatrix{*+[o][F-]{} \ar@{-}[rr] && *+[o][F-]{} \\
*+[F-]{}\ar@{-}[u]&&*+[F-]{}\ar@{-}[u]}
\end{gathered}
\ee
Notice that the contribution from a sector with vanishing magnetic charge, say $m_2=0$ is encoded in the dimension
\be
\Delta(m_1,0) = | m_1 |
\ee
That is the charge associated to the graph
\be
\xymatrix{*+[o][F-]{} \ar@{-}[rr]^2 && *+[F-]{} }
\ee
The dressing that corresponds to the two classical fields given by the two $U(1)$'s is $(1-t)^{-2}$. The refined Hilbert series then has the expression
\be\label{zwrgl}
H_{2,1}(t\,; z_1,z_2) = \frac{1}{(1-t)^2} \Big(1 + h^{(1)}_2 (t\,;z_1) + h^{(1)}_2(t\,;z_2) + h^{(2)}_1(t\,;z_1,z_2)\Big)
\ee
We have only to compute $h^{(2)}_1$, to obtain the following result
\be\label{H21ref}
H_{2,1}(t\,;z_1,z_2) =  \frac{1 + 2 t - \left(\frac{1}{z_1} + z_1+ \frac{1}{z_2}  + z_2 + \frac{1}{z_1z_2} + z_1 z_2\right) t^2 + 2 t^3 + t^4}{(1 - \frac{1}{z_1} t) (1 - t \, z_1) (1- \frac{1}{z_2} \, t) (1 - t \, z_2) (1-\frac{1}{z_1 z_2} t) (1 - t \, z_1 z_2)}
\ee
By multiplying and dividing the above expression by $(1-t)^2$, one finds that it can be rewritten in terms of $SU(3)$ characters:\footnote{ The same remark of footnote \ref{caveat} applies here. $\text{PE}[f(x)] \equiv \text{exp}\left(\sum_{n\geq 1} \frac{f(x^n)-f(0)}{n}\right)$.}
\be\label{H21ref}
H_{2,1}(t\,;z_1,z_2) = \text{PE}[ [1,1] \,t ] \times( 1-(1+[1,1])\, t^2 + 2 \, [1,1] \, t^3 - (1+[1,1])\, t^4 + t^6)
\ee
Indeed, the whole spectrum factor nicely in the $\text{Adj}^m$ characters of $SU(3)$:
\be\label{SU3char}
H_{2,1}(t\,;z_1,z_2) = \sum_{m=0}^\infty [m,m] \, t^{m}
\ee 
Consider now the expansion of the plethystic logarithm of the series at the lowest orders: One finds the following structure
\bea\label{PL21}
\text{PL}[H_{2,1}(t\,;z_1,z_2)] =& \left( 2 + \frac{1}{z_1} + z_1 + \frac{1}{z_2} + z_2 + \frac{1}{z_1 z_2}  + z_1 z_2 \right) \, t \\ &- \left( 3 + \frac{1}{z_1} + z_1 + \frac{1}{z_2} + z_2 + \frac{1}{z_1 z_2}  + z_1 z_2 \right) t^2 + O(t^3)
\eea
One can see that the two classical scalars $\phi_1$ and $\phi_2$, merge together with the six BPS monopole operators $V_{1,0}$, $V_{0,1}$, $V_{1,1}$ and their PCT conjugates to form the 8 states of the adjoint representation of $SU(3)$: $\phi_i$ represents the element in the Cartan, while the $V$'s are in one to one correspondence with the roots. Let us call $\Xi$ the $SU(3)$ adjoint representation so obtained. According to the Gaiotto--Witten criterion (see section 2.4.3 of \cite{GW3d}), the model has an $SU(3)$ flavor symmetry because we have a sequence of 2 balanced nodes. At level $t^2$ there is a relation. With respect to the $SU(3)$ weights, the relation is in the representation $[0,0]\oplus[1,1]$. There is a unique way of getting a relation of dimension $2$ compatible with the charges in eqn.\eqref{PL21} and with Bose statistics, that preserves the $SU(3)$ flavor symmetry. This fixes the Coulomb moduli space to be
\be\label{moduliA2p}
\text{Sym}^2 \, \Xi \Big|_{[0,0]\oplus[1,1]} = 0.
\ee
And indeed the structure of the expansion in eqn.\eqref{SU3char} confirms that these relations are enough to obtain the right multiplicities in the chiral ring. Notice that the corresponding moduli space is not a complete intersection, even if we have stopped the expansion of the plethystic logarithm at $O(t^3)$, it goes on indefinitely. However, the (complex) dimension of the moduli space is captured by the order of the highest pole at $t=1$ of the unrefined series. The unrefined series is
\be
H_{2,1}(t) = \frac{1 + 4 t  + t^2}{(1-t)^4} = \frac{6}{(t - 1)^4} + \frac{6}{(t - 1)^3} + \frac{1}{(t - 1)^2}
\ee
and the complex dimension of the moduli space is 4 as expected.

\bigskip

\noindent{\bf A 4d explanation of the $SU(3)$ symmetry.} The fact that the model $\ca_{2,1}$ has an $SU(3)$ flavor symmetry is not a mystery from the $4d$ $\cn=2$ perspective too. This follows from the fact that the singular Calabi--Yau hypersurface in this case is
\be\label{A22}
A_3(X_1,X_2) + A_3(Y_1,Y_2) = X_1^3 + X_2^2 + Y_1^3+Y_2^2 = 0
\ee
The singularity $X_1^3 + Y_1^3$ is well known to be equivalent to the $D_4$ singularity.\footnote{ To see this explicitly, it is sufficient to notice that the operator $X_1Y_2^2$ has $q(X_1Y_2^2) = 1$ and therefore it is an exactly marginal deformation for the corresponding $2d$ $\cn=(2,2)$ LG model.} This is mirrored by the fact that the two BPS quivers $A_2\boxtimes A_2$ and $D_4$ are mutation equivalent, and therefore, describe the same $4d$ $\cn=2$ model \cite{CV11}. The enhancement of the flavor symmetry to $SU(3)$ in this case follows if one identifies the $SU(3)$ Weyl group action on the charge lattice with the $D_4$  triality group. In a sense, our result completes what was obtained about this model in reference \cite{AMT}.

\bigskip

\noindent{\bf The $p>1$ case.}  Let us consider the refined Hilbert series for $p>1$. From eqn.\eqref{piden}, we obtain
\be
H_{2,p}(t\,;z_1,z_2) = \left(\frac{1-t^p}{1-t}\right)^2 \sum_{m=0}^\infty [m,m] \, t^{m p}
\ee
Again, let us consider the plethystic logarithm of it:
\bea
\text{PL}[H_{2,p}(t\,;z_1,z_2)] &= 2 t + \left(\frac{1}{z_1} + z_1 + \frac{1}{z_2} + z_2 + \frac{1}{z_1 z_2}  + z_1 z_2 \right) t^p\\& + \left( 3 + \frac{1}{z_1} + z_1 + \frac{1}{z_2} + z_2 + \frac{1}{z_1 z_2}  + z_1 z_2 \right) t^{2p} + \text{higher orders}
\eea
as one can easily see, we have the two classical scalars $\phi_1$ and $\phi_2$ with dimension 1, the six monopole operators $V_{1,0}$, $V_{0,1}$,$V_{1,1}$, and their PCT conjugates (lifted to dimension $p$), and a relation at dimension $2p$ (with the same structure of the $p=1$ case). Let us stress that from this expansion we see that there is no $SU(3)$ symmetry anymore: an $SU(3)$ flavor current would correspond to an adjoint representation at $\Delta=1$, this is there in the $p=1$ case, but it disappears here. This is consistent with the Gaiotto--Witten criterion: indeed from this criterion the flavor symmetry is $U(1)\times U(1)$. From the Hilbert series we can compute the dimension of moduli space, indeed, the unrefined series is such that
\be
H_{2,p}(t) = \frac{1 + 4 t^p + t^{2p}}{(1 - t)^2 (1 - t^p)^2} \sim  \frac{6}{p^2(t-1)^4} \qquad \text{ as } t \to 1
\ee
and the complex dimension of the moduli space is 4.

\bigskip

\noindent{\bf Remark.} The model $\ca_{2,2}$ corresponds to the BPS quiver $A_2\boxtimes A_5$, in the mutation class for the $E_8^{(1,1)}$ exceptional SCFT \cite{CV11}. This is one of the 11 exceptional models that does not have a known Gaiotto curve, while having a gauge group $SU(2)$ and being complete in the sense of \cite{CV11}. This model in four dimensions corresponds to a $SU(2)$ SYM sector weakly gauging a $SU(2)$ subgroup of the flavor symmetry of some Argyres--Douglas systems of type $D$, namely $D_5$ and $D_3$, coupled to a hypermultiplet in the fundamental. As we are going to see in the next section, also the $E_7^{(1,1)}$ SCFT belongs to the class of models analyzed in this paper.

\subsection{The Higgs branch of the systems of type $\ca_{3,p}$}
The system of type $\ca_{3,1}$ is completely described by the 3d model associated to the graph $\ca_{3,1}^1$. The corresponding effective magnetic charges are
\be
\Delta(m_1,m_2,m_3) = \frac{1}{2}\Big(|m_1-m_2| + |m_1-m_3| + |m_2-m_3| + |m_1|+ |m_2|+ |m_3|\Big)
\ee
Proceeding as before, we see that the contribution for sectors with 2 vanishing magnetic charges are computed by $h^{(1)}_3(t\,; z_i)$ functions, and contributions for sectors with 1 vanishing magnetic charge are encoded in $h^{(2)}_2(t \, ; z_i , z_j)$ functions. We are only left with the computation of the $h^{(3)}_1(t\,;z_1,z_2,z_3)$ function. The $p=1$ refined Hilbert series, is given by
\bea
H_{3,1}(t\,;z_1,z_2,z_3) =& \frac{1}{(1-t)^3}\Big(1 + h^{(1)}_3(t\,;z_1)+ h^{(1)}_3(t\,;z_2)+ h^{(1)}_3(t\,;z_3)\\
& + h^{(2)}_2(t\,;z_1,z_2)+ h^{(2)}_2(t\,;z_2,z_3)+ h^{(2)}_2(t\,;z_1,z_3)\\
& + h^{(3)}_1(t\,;z_1,z_2,z_3) \Big)
\eea
In this case however, the explicit expression of the refined Hilbert series is rather involved, and we prefer to write down only the unrefined Hilbert series, that is
\be
H_{3,1}(t) = \frac{1}{(t-1)^3} \frac{1 - t^{\frac{1}{2}} + t + 5 t^{\frac{3}{2}} + 5 t^3 + t^{\frac{7}{2}} - t^4 + t^{\frac{9}{2}}}{(t^{\frac{1}{2}}-1)^3 (1 + t) (1 + t^\frac{1}{2} + t)^2}
\ee
This can be rewritten in the following form
\be
H_{3,1}(t) =  \frac{1 - t^{\frac{1}{2}} + t + 5 t^{\frac{3}{2}} + 5 t^3 + t^{\frac{7}{2}} - t^4 + t^{\frac{9}{2}}}{ (1 - t^\frac{1}{2}) (1 - t)^2 (1 - t^\frac{3}{2})^2 (1 - t^2)}
\ee
In particular notice that
\be
H_{3,1}(t) \sim \frac{16}{3(1-t)^6} \qquad\text{ as } t\to 1
\ee
therefore, the moduli space has complex dimension 6 as expected. As before the plethystic logarithm has no finite order, and therefore the space is not a complete intersection. The first terms in the expansion of the plethystic logarithm of the refined Hilbert series are the following
\bea
\text{PL}[H_{3,1}(t\,;z_1,z_2,z_3)]=& \, 3 \,  t + \Big(\frac{1}{z_1} + z_1 + \frac{1}{z_2} + z_2 + \frac{1}{z_3}+ z_3  + \frac{1}{z_1 z_2 z_3} + z_1 z_2 z_3\Big) t^\frac{3}{2}\\
&+\Big(\frac{1}{z_1 z_2} + z_1 z_2 + \frac{1}{z_1 z_3} + z_1 z_3 + \frac{1}{z_2 z_3}  + z_2 z_3\Big)t^2\\
&- \Big(4 + \frac{2}{z_1 z_2} + 2 z_1 z_2 + \frac{2}{z_1 z_3} + 2 z_1 z_3 + \frac{2}{z_2 z_3}  + 2 z_2 z_3\Big) t^3 \\
&+ \dots 
\eea
We see that there are:
\begin{itemize}
\item 3 scalar operators with dimension $\Delta=1$ (the 3 classical $\phi_i$ $i=1,2,3$ associated to the 3 $U(1)$'s); 
\item 8 scalar operators with dimension $\Delta=3/2$ (the monopole operators $V_{1,0,0}$, $V_{0,1,0}$, $V_{0,0,1}$, $V_{1,1,1}$, and their PCT conjugates);
\item 6 scalar operators with dimension $\Delta=2$ (the monopole operators $V_{1,1,0}$, $V_{1,0,1}$, $V_{0,1,1}$);
\item 16 relations in between them with dimension $\Delta=3$.
\end{itemize}
Of course, the algebraic variety we are considering is not a complete intersection, in between the first 16 relations there are going to be syzygies. However, these should be enough to completely determine the moduli space as an algebraic variety. The structure of the relations are easily guessed using the  $\mathfrak{S}_3$ symmetry of the problem:
\bea\label{vague}
&V_{1,0,0} V_{0,1,0} = V_{1,1,0} \times \text{scalars}\qquad V_{-1,0,0} V_{1,1,1} = V_{0,1,1} \times \text{scalars}\\
&V_{1,0,0}V_{-1,0,0} = \text{scalars}\qquad\qquad\quad V_{1,1,1}V_{-1,-1,-1} = \text{scalars}\\
& + \mathfrak{S}_3 \text{ symmetric permutations }\\
& +  \text{PCT conjugates}
\eea
The plethystic Log of the refined Hilbert series, however, is not powerful enough to constrain the structure of the scalar contributions.

\bigskip

\begin{figure}
\begin{center}
\begin{tabular}{cc}
\includegraphics[width=0.4\textwidth,height=0.4\textwidth]{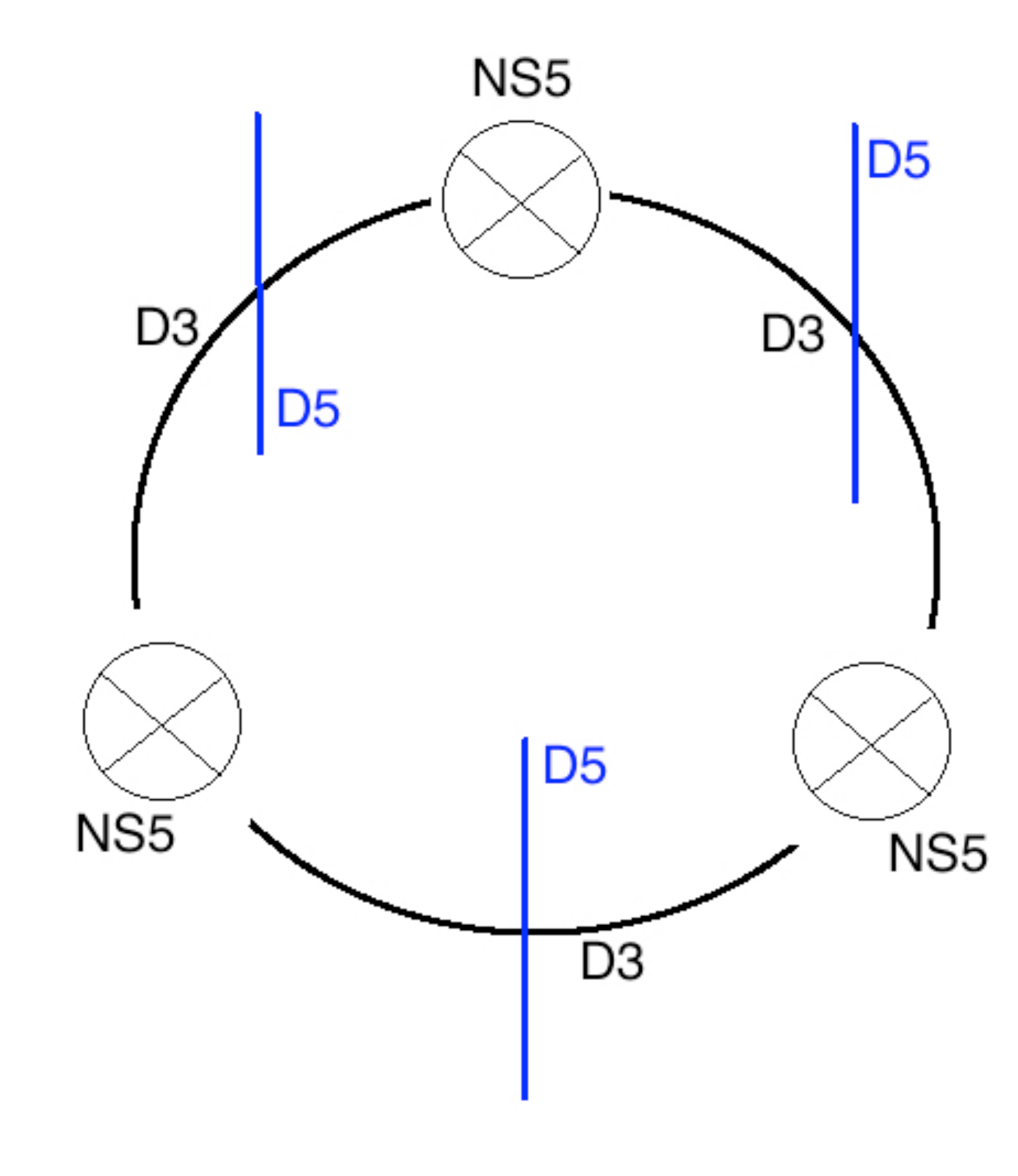}&
$\begin{gathered}\xymatrix{*+[F-]{}&&*+[F-]{}\\*+[o][F-]{} \ar@{-}[u]\ar@{-}[rr]&&*+[o][F-]{} \ar@{-}[u]\\&*+[o][F-]{}\ar@{-}[ur]\ar@{-}[ul]&\\&*+[F-]{} \ar@{-}[u]&\\}\end{gathered}$\\
\end{tabular}
\end{center}
\caption{The graph $\ca^1_{3,1}$ and the corresponding brane network.}\label{thefig}
\end{figure}

For this special case, however, the problem can be overcome. The graph $\ca^1_{3,1}$ is drawn in figure \ref{thefig} together with the system of NS5 - D5 - D3 branes that engineers the corresponding theory. Mirror symmetry at the level of the brane system switches NS5 branes with D5 branes: one can easily convince himself that the system is self--mirror!! From this fact it follows that the Higgs branch of the 4d $\cn=2$ model $\ca_{3,1}$ can be identified with the one of the $\ca_{3,1}^1$ model. Let us determine it.

\bigskip

\begin{table}

\end{table}

The quiver relevant for determining the Higgs branch is the following
\be
\begin{gathered}
\xymatrix{&*++[F-]{}\ar@/_0.5pc/[d]_{Q_1}&&&&*++[F-]{}\ar@/_0.5pc/[d]_{Q_2}\\
&*++[o][F-]{}  \ar@(ul,dl)[]_{\Phi_1} \ar@/_0.7pc/[u]_{\widetilde{Q}_1}\ar@/^0.7pc/[rrrr]^{H_3}\ar@/^0.7pc/[ddrr]^{\widetilde{H}_2}&&&&*++[o][F-]{}\ar@/^0.5pc/[llll]^{\widetilde{H}_3} \ar@/_0.7pc/[u]_{\widetilde{Q}_2}\ar@/^0.7pc/[ddll]^{H_1}\ar@(ur,dr)[]^{\Phi_2}&\\
&&&&\\
&&&*++[o][F-]{}\ar@/^0.5pc/[uull]^{H_2}\ar@/^0.5pc/[uurr]^{\widetilde{H}_1}\ar@/^0.5pc/[dr]^{\widetilde{Q}_3}[u]\ar@(ul,dl)[]_{\Phi_3}&\\
&&&&*++[F-]{} \ar@/^0.5pc/[ul]^{Q_3}[u]&\\}
\end{gathered}
\ee
Let us define the following gauge invariant operators $M_i = Q_i\widetilde{Q}_i$, $L_i = H_i\widetilde{H}_i$. The relevant F--terms are (here and always below $i=1,2,3$ and is taken mod 3):
\be
M_i = L_{i+1} + L_{i+2}
\ee
The gauge invariant operators with $\Delta\leq2$ are described below
\begin{center}
\begin{tabular}{|ccc|}
\hline
$\Delta$&Operator &Symbol\\
$1$&$H_i\widetilde{H}_i$&$L_i$\\
$3/2$&$Q_i H_{i+2}\widetilde{Q}_{i+1}$&$A_i$\\
$3/2$&$H_1H_2H_3$&$W$\\
$2$&$Q_i H_{i-1}H_i\widetilde{Q}_{i-1}$& $B_i$\\
\hline
\end{tabular}
\end{center}
One can easily verify that the only relations that these operators satisfy at $\Delta = 3$ are the following ones
\be
\begin{aligned}
&A_i A_{i+1} = B_i \, (L_{i+2} + L_{i+3}) \quad \text{and PCT conjugate} \\
&A_i \widetilde{W} = \widetilde{B}_{i+1} \, L_{i+2} \quad \text{and PCT conjugate} \\
&W \widetilde{W} = L_1 \, L_2 \, L_3 \\
& A_i \widetilde{A}_i = (L_{i+1} + L_{i+2})(L_i+L_{i+2}) L_{i+2}\\
\end{aligned}
\ee
where again $i$ is taken mod 3. These relations clearly matches the plethystic Log of the refined Hilbert series, if we identify
\be
\begin{gathered}
V_{1,0,0} = A_1\qquad V_{0,1,0} = A_2 \qquad V_{0,0,1} = A_3\\
V_{1,1,0} = B_1\qquad V_{0,1,1} = B_2 \qquad V_{1,0,1} = B_3\\
V_{1,1,1} = W
\end{gathered}
\ee

\bigskip

\noindent{\bf Exceptional $E^{(1,1)}_7$ complete SCFT.} The model $\ca_{3,1}$ has BPS quiver in the mutation class of $A_3\boxtimes A_3$: this mutation class corresponds to the exceptional complete theory $E^{(1,1)}_7$. This is a very interesting $\cn=2$ SCFT: as its higher rank cousin we have discussed in the remark at the end of the previous section, this system  belongs to the set of the 11 exceptional $4d$ $\cn=2$ complete models, and it describes a $SU(2)$ SYM sector weakly gauging a $SU(2)$ subgroup of the flavor symmetry of two Argyres--Douglas systems of type $D_4$ together with a quark in the fundamental. 

\bigskip

\noindent {\bf Comments about the $p>1$ case.} For $p>1$, we have from eqn.\eqref{piden}
\be
H_{3,p}(t) = \frac{1}{(1-t)^3} \frac{1 - t^{\frac{p}{2}} + t^p + 5 t^{\frac{3p}{2}} + 5 t^{3p} + t^{\frac{7p}{2}} - t^{4p} + t^{\frac{9p}{2}}}{(t^{\frac{p}{2}}-1)^3 (1 + t^p) (1 + t^\frac{p}{2} + t^p)^2}
\ee
or, alternatively
\be
H_{3,p}(t) =  \left( \frac{1-t^p}{1-t} \right)^3  \frac{1 - t^{\frac{p}{2}} + t^p + 5 t^{\frac{3p}{2}} + 5 t^{3p} + t^{\frac{7p}{2}} - t^{4p} + t^{\frac{9p}{2}}}{ (1 - t^\frac{p}{2}) (1 - t^p)^2 (1 - t^\frac{3p}{2})^2 (1 - t^{2p})}
\ee
where again we write only our expression for $H_{3,p}(t)$ for brevity. This is enough to check that
\be
H_{3,p}(t) \sim \frac{16}{3 p^3 (1-t)^6} \qquad \text{ for } t \to 1
\ee
as expected.

The plethystic logarithm for $p>1$ has the same structure of the $p=1$ case: we obtain
\bea
\text{PL}[H_{3,p}(t\,;z_1,z_2,z_3)]=& \, 3 \,  t + \Big(\frac{1}{z_1} + z_1 + \frac{1}{z_2} + z_2 + \frac{1}{z_3}+ z_3  + \frac{1}{z_1 z_2 z_3} + z_1 z_2 z_3\Big) t^\frac{3 p}{2}\\
&+\Big(\frac{1}{z_1 z_2} + z_1 z_2 + \frac{1}{z_1 z_3} + z_1 z_3 + \frac{1}{z_2 z_3}  + z_2 z_3\Big)t^{2p}\\
&- \Big(4 + \frac{2}{z_1 z_2} + 2 z_1 z_2 + \frac{2}{z_1 z_3} + 2 z_1 z_3 + \frac{2}{z_2 z_3}  + 2 z_2 z_3\Big) t^{3p} \\
&+ \dots 
\eea
The monopole operators are lifted in dimensions $3p/2$ and $2p$, while the relation involves dimension $3p$ operators.

\bigskip

\section*{Ackowledgements}
We warmly acknowledge Hiraku Nakajima, Francesco Benini, and Dan Xie for very useful discussions. We thank the organizers of the October 2013 workshop ``Quiver Varieties'' at the Simons Center in Stony Brook, where this project has been conceived. The work of MDZ is supported by the NSF grant PHY-1067976.

\end{document}